\documentstyle[psfig,epsfig]{mn}

\newif\ifAMStwofonts

\def\simlt{\lower.5ex\hbox{$\; \buildrel < \over \sim \;$}}
\def\simgt{\lower.5ex\hbox{$\; \buildrel > \over \sim \;$}}

\ifoldfss
  \ifCUPmtlplainloaded \else
    \NewTextAlphabet{textbfit} {cmbxti10} {}
    \NewTextAlphabet{textbfss} {cmssbx10} {}
    \NewMathAlphabet{mathbfit} {cmbxti10} {} 
    \NewMathAlphabet{mathbfss} {cmssbx10} {} 
  \fi
  \ifAMStwofonts
    \ifCUPmtlplainloaded \else
      \NewSymbolFont{upmath} {eurm10}
      \NewSymbolFont{AMSa} {msam10}
      \NewMathSymbol{\upi}     {0}{upmath}{19}
      \NewMathSymbol{\umu}     {0}{upmath}{16}
      \NewMathSymbol{\upartial}{0}{upmath}{40}
      \NewMathSymbol{\leqslant}{3}{AMSa}{36}
      \NewMathSymbol{\geqslant}{3}{AMSa}{3E}

    \fi
  \fi
\fi 

\ifnfssone
  \newmathalphabet{\mathit}
  \addtoversion{normal}{\mathit}{cmr}{m}{it}
  \addtoversion{bold}{\mathit}{cmr}{bx}{it}
  \newmathalphabet{\mathbfit} 
  \addtoversion{normal}{\mathbfit}{cmr}{bx}{it}
  \addtoversion{bold}{\mathbfit}{cmr}{bx}{it}
  \newmathalphabet{\mathbfss} 
  \addtoversion{normal}{\mathbfss}{cmss}{bx}{n}
  \addtoversion{bold}{\mathbfss}{cmss}{bx}{n}
  \ifAMStwofonts
    \ifCUPmtlplainloaded \else
      \UseAMStwoboldmath
      \makeatletter
      \new@mathgroup\upmath@group
      \define@mathgroup\mv@normal\upmath@group{eur}{m}{n}
      \define@mathgroup\mv@bold\upmath@group{eur}{b}{n}
      \edef\UPM{\hexnumber\upmath@group}
      \new@mathgroup\amsa@group
      \define@mathgroup\mv@normal\amsa@group{msa}{m}{n}
      \define@mathgroup\mv@bold\amsa@group{msa}{m}{n}
      \edef\AMSa{\hexnumber\amsa@group}
      \makeatother
      \mathchardef\upi="0\UPM19
      \mathchardef\umu="0\UPM16
      \mathchardef\upartial="0\UPM40
      \mathchardef\leqslant="3\AMSa36
      \mathchardef\geqslant="3\AMSa3E
    \fi
  \fi
\fi 

\ifnfsstwo
  \DeclareMathAlphabet{\mathbfit}{OT1}{cmr}{bx}{it}
  \SetMathAlphabet\mathbfit{bold}{OT1}{cmr}{bx}{it}
  \DeclareMathAlphabet{\mathbfss}{OT1}{cmss}{bx}{n}
  \SetMathAlphabet\mathbfss{bold}{OT1}{cmss}{bx}{n}
  \ifAMStwofonts
    \ifCUPmtlplainloaded \else
      \DeclareSymbolFont{UPM}{U}{eur}{m}{n}
      \SetSymbolFont{UPM}{bold}{U}{eur}{b}{n}
      \DeclareSymbolFont{AMSa}{U}{msa}{m}{n}
      \DeclareMathSymbol{\upi}{0}{UPM}{"19}
      \DeclareMathSymbol{\umu}{0}{UPM}{"16}
      \DeclareMathSymbol{\upartial}{0}{UPM}{"40}
      \DeclareMathSymbol{\leqslant}{3}{AMSa}{"36}
      \DeclareMathSymbol{\geqslant}{3}{AMSa}{"3E}
    \fi
  \fi
\fi 

\ifCUPmtlplainloaded \else
  \ifAMStwofonts \else 
    \def\upi{\pi}
    \def\umu{\mu}
    \def\upartial{\partial}
  \fi
\fi

\title{Joint Cosmic Shear Measurements with the Keck and William Herschel
Telescopes}
\author[D. J. Bacon et al.]
{David~J.~Bacon,$^{1,2,3}$\thanks{E-mail: djb@roe.ac.uk}
Richard~J.~Massey,$^3$ Alexandre~R.~Refregier$^{1,3}$ \& Richard~S.~Ellis$^1$ \\
$^1$ California Institute of Technology, Pasadena CA 91125, USA\\
$^2$ Institute for Astronomy, Blackford Hill, Edinburgh EH9 3HJ, UK\\
$^3$ Institute of Astronomy, Madingley Road, Cambridge CB3 OHA, UK \\}

\date{Accepted ---. Received ---; in original form ---.}

\pagerange{\pageref{firstpage}--\pageref{lastpage}}
\pubyear{2002}

\begin{document}

\maketitle

\label{firstpage}

\begin{abstract} The recent measurements of weak lensing by
large-scale structure present significant new opportunities for
studies of the matter distribution in the universe. Here, we present a
new cosmic shear survey carried out with the Echelle Spectrograph and
Imager on the Keck II telescope. This covers a total of 0.6 square
degrees in 173 fields probing independent lines of sight, hence
minimising the impact of sample variance. We also extend our
measurements of cosmic shear with the William Herschel Telescope
(Bacon, Refregier \& Ellis 2000) to a survey area of 1 square degree.
The joint measurements with two independent telescopes allow us to
assess the impact of instrument-specific systematics, one of the major
difficulties in cosmic shear measurements. For both surveys, we
carefully account for effects such as smearing by the point spread
function and shearing due to telescope optics. We find negligible
residuals in both cases and recover mutually consistent cosmic shear
signals, significant at the $5.1\sigma$ level. We present a simple
method to compute the statistical error in the shear correlation
function, including non-gaussian sample variance and the covariance
between different angular bins.  We measure shear correlation
functions for all fields and use these to ascertain the amplitude of
the matter power spectrum, finding $\sigma_{8} \left(
\frac{\Omega_{m}}{0.3} \right)^{0.68} = 0.97 \pm 0.13$ with
$0.14<\Omega_m<0.65$ in a $\Lambda$CDM model with $\Gamma=0.21$. These
$68\%$ CL uncertainties include sample variance, statistical noise,
redshift uncertainty, and the error in the shear measurement
method. The results from our two independent surveys are both
consistent with measurements of cosmic shear from other groups. We
discuss how our results compare with current normalisation from
cluster abundance.
\end{abstract}


\section{Introduction} \label{intro}

Weak gravitational lensing of background galaxies by intervening
large-scale structure (`cosmic shear') provides direct information
about the total mass distribution in the universe, regardless of its
nature and state. A measurement of cosmic shear thus bridges the gap
between theory, which is primarily concerned with dark matter, and
observation, which generally probes only luminous matter. The recent
measurements of coherent distortion of faint galaxies by several
groups (van Waerbeke et al. 2000; Bacon, Refregier \& Ellis 2000
[BRE]; Wittman et al. 2000; Kaiser, Wilson \& Luppino 2000; Maoli et
al.  2001; Rhodes, Refregier \& Groth 2001; van Waerbeke et al 2001;
H\"{a}mmerle et al.  2001; Hoekstra et al. 2002; Refregier, Rhodes \&
Groth 2002) has triggered great interest in the provision of new
constraints on the amount and distribution of dark matter, together
with measurements of several cosmological parameters.

If intrinsic galaxy orientations are essentially random in a given
survey (which requires the survey to be sufficiently deep; see Brown
et al. 2000; Catelan et al. 2000; Heavens et al 2001; Croft \& Metzler
2001; Crittenden et al 2001), any coherent alignment must arise from
distortion due to weak lensing. Light paths from galaxies projected
close together on the sky pass through, and are gravitationally
distorted by, the same dark matter concentrations. This coherent
distortion contains valuable cosmological information (eg. Bernardeau
et al. 1997; Jain \& Seljak 1997; Kamionkowski et al. 1997; Kaiser
1998; Hu \& Tegmark 1998). In particular, the variance of the
distortion field measures the amplitude of density fluctuations ($\sim
\sigma_8 \Omega_m^{0.5}$). This shear measurement is free from
assumptions about gaussianity or the $M-T$ relation, and whilst the
shear-based measurement is currently comparable in precision to that
from local cluster abundance, further progress is limited solely by
the number of fields observed.

The validity of results from cosmic shear surveys depends sensitively
on the treatment of systematic errors. A further issue arises from
sample (or `cosmic') variance, the impact of which can be limited by
using numerous independent sightlines to complement panoramic imaging
of a few selected areas. With these motivations in mind, we present a
comparison of the cosmic shear observed with two independent
instruments (Keck and WHT), using two different survey strategies.

In this paper, we describe the first cosmic shear measurements with
the $8' \times 2'$ Echelle Spectrograph and Imager (ESI) on the Keck
II telescope. This Keck survey reaches a depth of $z \simeq 1.0$,
comparable to other recent cosmic shear surveys (e.g. van Waerbeke et
al 2001, Bacon et al 2000). However, the much faster acquisition of
fields with ESI in comparison to 4m telescopes such as the William
Herschel Telescope allows us to obtain very many more fields
(173 in the final survey). This improves the cosmic
shear signal measured, by minimising the contribution to noise of the
sample variance, i.e. the error upon the mean lensing signal resulting
from the measurement of shear on only a limited number of lines of
sight. In order to measure the cosmic shear signal, we analyse the
correlation function of the shear on various scales, and obtain
constraints on cosmological parameters using a $\chi^2$ fit to
theoretical predictions upon varying these parameters.

In addition to this investigation, we extend our original detection of
cosmic shear on the 4.2m WHT (in BRE), to a
measurement of the correlation function of the distortion field for
this dataset. A more precise measurement is afforded by the increase
in the number of WHT fields to a total of 20, with a further increase
in area due to the larger $16'\times16'$ size of field with the new
WHT mosaic camera.

This paper is organised as follows. In \S\ref{strategy} we discuss our
observational strategy for measuring cosmic shear with both
telescopes. We then describe (\S\ref{keck}) the procedure for
observing with the Keck telescope, and explain the method used for
reducing the data. We follow this by a treatment of the systematic
effects associated with the Keck data, such as shear distortion from
the camera and the anisotropic smear due to telescope tracking.

Having dealt with the Keck data up to the stage of measuring galaxy
shapes, we review the procedure for obtaining the WHT data in
\S\ref{wht}, followed by the approach adopted for data reduction and
removal of systematic effects.

We proceed to develop the correlation function formalism in
\S\ref{cosshr}, which we use for precision measurements of the cosmic
shear, in \S\ref{results}. We then interpret the shear signal in terms
of cosmological models in order to obtain limits on the cosmological
parameters $\sigma_8$ and $\Omega_m$.  Our conclusions are summarised
in \S\ref{conclusions}.

\section{Survey Strategy}  \label{strategy}

The aim of our Keck and WHT surveys has been to acquire deep
($z\simeq 1$) fields representing numerous independent lines of
sight, sufficiently scattered to sample independent structures and
thus minimise uncertainties due to sample variance. These lines of
sight must be chosen in a quasi-random fashion, without regard to the
presence or absence of mass concentrations, in order to obtain a
representative sample of the mass fluctuations in the universe. Here
we describe the strategy adopted for the two surveys, based on that
in BRE.

The survey fields for both Keck and WHT were selected by choosing a
sparse ($>2^\circ$ separation for statistical independence)
grid of coordinates spanning the range accessible to the telescope at
a given time. We tuned the Galactic latitude of the grid to ensure
$\sim$50 unsaturated stars within the Keck field of view, and
$\sim$200 in the WHT fields, so that the anisotropic PSF and the
camera distortion could be carefully mapped (see below). The STScI
Digitised Sky Survey was then used to find an appropriate final field
near each set of coordinates, avoiding stars brighter than $R<11$ in
the APM and GCC catalogues, to prevent large areas of saturation or
ghost images.

\begin{figure} \epsfig{figure=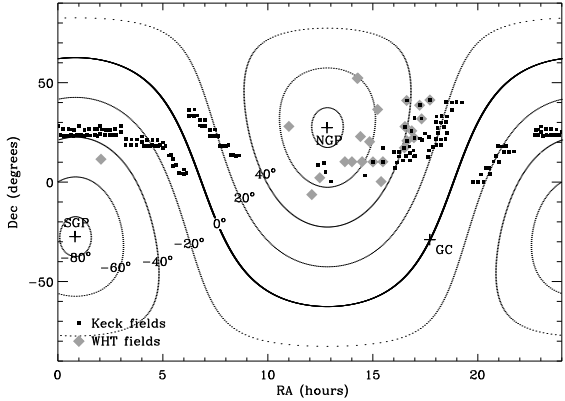,width=80mm} \caption{The sky
location of the cosmic shear fields. Galactic latitudes of $0^\circ$,
$\pm 20^\circ$, $\pm 40^\circ$, $\pm 60^\circ$ are shown as contours;
the Galactic centre and poles are shown as a cross.}
\label{fig:skymap} \end{figure}

As a final constraint, each field was observed within $20^\circ$ of
zenith for both telescopes. This minimises smearing due to
atmospheric refraction for Keck, which does not have an Atmospheric
Dispersion Corrector (see section \ref{results} and figure
\ref{fig:stargal} for confirmation that this is not a limiting
systematic). For both telescopes the constraint minimises any image
distortion associated with telescope or instrument flexure. Figure
\ref{fig:skymap} shows the positions of the resulting selected survey
fields on the sky.

We must now determine the depth to which to observe these fields. We
have shown in BRE that the cosmic shear signal is measurable with WHT
images having a 1 hour exposure length, corresponding to a usable
galaxy sample with  median source redshift $z_{s}=0.8$. We have
further demonstrated in Bacon et al (2001) that longer exposures do
not improve the signal greatly, as beyond this depth galaxy shapes
are seriously degraded by a typical ground-based PSF. On the other
hand, shorter exposures face the danger of considerable contamination
by intrinsic alignments of galaxies (e.g. Heavens et al 2001). We
therefore aimed to probe the same redshift range with our new Keck
survey. An exposure time of 10 minutes
was calculated to allow $5\sigma$ detections of point sources at $R=26$, given
the optics throughput, sky background in the Keck $R$ filter
frequency range and the quantum efficiency of the ESI CCD. However,
as we shall see, the better seeing during our observations with Keck
results in a slightly fainter magnitude limit than the WHT survey;
this is entirely acceptable, as we can compare results by scaling the
predictions according to the equation $\sigma_\gamma \propto z^{0.8}$
(see e.g. BRE).

\begin{figure} \epsfig{figure=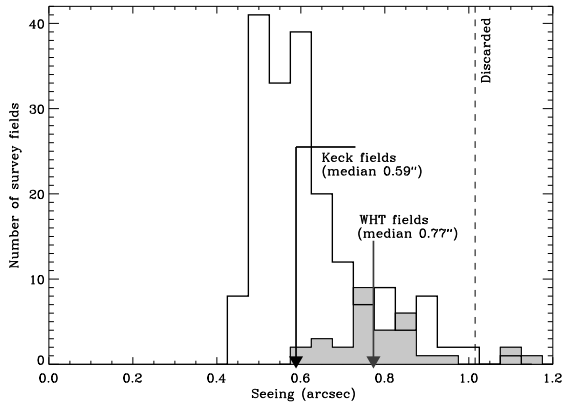,width=80mm}
\caption{Histogram of seeing (FWHM) in all survey shear fields after
stacking dithered exposures. Keck data are shown in white; WHT data
are shown as shaded. Each $8' \times 16'$ WHT chip is counted
separately. Combining survey area from the two telescopes, the median
seeing is 0.73''; no data with seeing worse than 1'' is used.}
\label{fig:seeing} \end{figure}

\begin{figure} \epsfig{figure=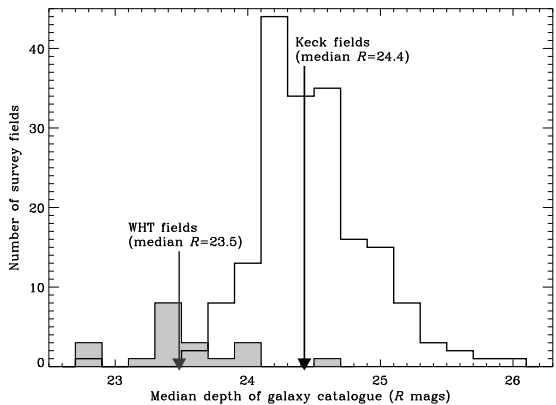,width=80mm} \caption{Median
reddening-corrected $R$-band magnitude of the galaxies in each field
that are used in the final object catalogue. The median depth of the
combined surveys is $R$=24.0.} \label{fig:depth} \end{figure}

The fact that gravitational lensing is achromatic permits us a free
choice of photometric band for our observations. However, we should
note that $R$ and $I$ afford the most efficient deep imaging in a
given exposure time. Due to fringing in the $I$ band with the EEV CCD
in use at the WHT, we choose to image in $R$ with both telescopes. The
Keck $R$-band is a specially constructed filter with similar
throughput and spectral range to the WHT Harris $R$; the slightly
different galaxy distribution probed by this filter is taken into
account later in our redshift uncertainty estimates.

\section{Keck data} \label{keck}
\subsection{Observations}

As we briefly mentioned above, the great advantage which Keck
presents in measuring cosmic shear is its speed in achieving the
necessary depth; 10 minute exposures reach $R=26$ (contrast the 1
hour necessary for such a depth with WHT). We can therefore obtain
very many independent lines of sight per night, thus reducing the
impact of sample variance.

We observed 173 $2'\times8'$ ESI fields using a specially made $R$
filter ($\bar{\lambda}=6657$\AA, effective FWHM$(\lambda)=1200$\AA),
over the course of 6 nights in June 2000, November 2000 and May 2001.
The necessary imaging observations were done as a good seeing
override on an independent spectroscopic programme. The pixel size of
$0.153''$ for ESI is considerably finer than that for WHT, allowing
better sampling of the galaxy images.

For each of our fields, three separate exposures were taken, each
offset by $5''$. This enables the continual re-calibration of optical
distortions in the telescope and camera (see section \ref{keckdist}),
and the removal of cosmetic defects and cosmic rays. All the fields
were observed as they passed near the meridian (in order to minimise
image distortion from the atmosphere and from telescope or instrument
flexure) but no closer than 5$^\circ$ to zenith (to minimise sky
rotation and potential tracking errors on this alt-az telescope).
Bias frames, sky flats and dome flats were acquired at the start and
end of each night, and standard star observations were taken regularly
throughout each night. The telescope was focused several times per
night to minimise camera distortions.

As exposures were observed, we used our real-time SExtractor-based
(Bertin \& Arnouts 1996) software on recently completed exposures in
order to monitor PSF size and stellar ellipticity. Figure
\ref{fig:seeing} records the seeing values for all fields; we found
that the median seeing for the observations was $0.59''$, with 75\% in
seeing better than $0.7''$. As we discuss below, this excellent
quality data yields a low level of noise on the estimates of cosmic
shear.

Furthermore, the rms stellar ellipticity for the images was
$\sigma_{e*}=0.035$. This relatively low value facilitates the
necessary PSF corrections (see e.g. Bacon et al 2001, Erben et al
2001). The stellar ellipticity on Keck fields is often found to be due
to tracking errors, seen as a uniform PSF anisotropy across the field
of view (see figure \ref{fig:kecke} for an example and discussion in
$\S$\ref{psf}). In exposures with excellent tracking, the dominant
effect appears to be astigmatism due to the fact that the CCD is
slightly skewed with respect to the focal plane, thus probing optical
conditions slightly above and below the focus.

In terms of depth, with magnitudes calibrated using photometric
standards, we found that the median magnitude of all galaxies detected
was $R$=25.1, with an {\tt imcat} signal-to-noise of 5.0 being reached
for galaxies at $R$=25.8. We keep $\simeq 27.5$ resolved
galaxies per square arcminute in our final object catalogue (see
section \ref{psf} for the selection criteria), which corresponds to a
final magnitude median of $R\simeq24.4 \pm 0.2$, including reddening
corrections. The distribution of magnitudes for the final galaxy
catalogue extends substantially fainter than this median, with the
galaxy count as a function of magnitude dropping to 50\% of maximum at
$R=25.2$. According to e.g. Cohen et al (2000), our median magnitude
of $R=24.4$ corresponds to a median source redshift of $z_s \simeq
1.0$. We can estimate the error on this by taking Cohen et al's
measured interquartile range of redshifts for galaxies at this
magnitude, $IQ$=1.6-0.55=1.05, and using the estimate of the error on
the mean for a gaussian distribution, $\sigma=IQ / (1.35 \sqrt{n}) =
0.1$, where the number of galaxies with redshifts measured
$n=49$. Thus we will use an estimate of the median redshift $z_s = 1.0
\pm 0.1$. This quantity may be subject to additional sample variance,
which would increase the error; we will therefore quote our redshift
error separately to other statistical errors for comparison with
future redshift surveys.

\subsection{Data Reduction} \label{keckdr}

These deep images were reduced following standard methods. Bias
subtraction of the science exposures and sky-flats was individually
calibrated for each image using the overscan regions at the edge of
the chip. The science exposures were then trimmed and divided by a
median composite (with 3$\sigma$ clipping) of the flat fields for a
given night. To eliminate any remaining background gradients, the
science images were also divided by a stack of all the
(median-normalised) exposures from that night. In contrast to the WHT
data, no fringing is observed on the ESI images, due to a thicker CCD
than that for the WHT camera; this simplifies the reduction
considerably.

The multiple exposures for each field were aligned by cross-matching
common objects in SExtractor catalogues (Bertin \& Arnouts 1996).
Typically $\simeq$250 objects were found in common per Keck field,
thus providing a stringent estimate of the mean dither offset (to
0.02 pixel accuracy). The exposures were then shifted by the required
non-integer number of pixels via a linear interpolation using the
IRAF {\tt imshift} routine. As we discuss below in \S\ref{keckdist},
rotations between dithers, and astrometric distortions due to the
telescope and camera optics, are found to be negligible in our
analysis.

The resulting registered exposures were then divided by their median
values to normalise their background levels and averaged with
3$\sigma$ clipping to remove cosmic rays and cosmetic defects.
Remaining problematic areas (edges, bad columns, regions containing
light leaks from stars outside the field of view, spikes from slightly
saturated stars, and a $<1'$ square unresponsive region in the corner
of the CCD) were flagged and are not used in our cosmic shear analysis
(see also $\S$\ref{keckmask}). Note that saturation spikes are also
largely excluded from our object catalogues by an initial $e < 0.5$
cut, but a local sky background gradient required that a few galaxies
within 2'' of bright stars also be discarded. Fortunately with the
small field of ESI we can selectively avoid most stars which are
bright enough to saturate. After these maskings, a reliable $\sim$12.8
square arcminutes per survey field remains.

Having obtained carefully reduced data, the next step is to catalogue
the shapes of galaxies and to estimate their gravitational shear. In
order to do this, we must correct for any shear introduced by the
telescope itself, together with tracking errors and atmospheric
smearing which convolve the galaxy shapes: these effects can mimic a
coherent distortion which must be removed. We first turn to the issue
of instrumental optical distortions, mimicking shear.

\subsection{Instrumental Distortions} \label{keckdist}

\begin{figure} 
\hspace{0.5in}
\psfig{figure=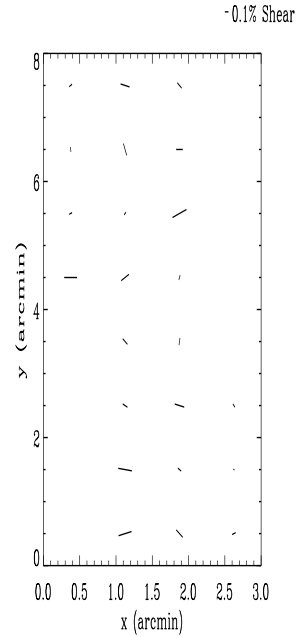,width=40mm,height=100mm}
\caption{Example instrumental shear pattern for ESI. Each bar
represents the magnitude and orientation of astrometric distortions
averaged over 20 sets of three dithered exposures. The illuminated
area of the CCD is rotated from the $x-y$ axes by $7^\circ$,
accounting for the occupied bins seen here.}
\label{fig:keckdist} \end{figure}

Our method for determining the shear field induced by the telescope
and camera, using the offsets in SExtractor catalogues of our dithered
astrometric frames, is fully documented in BRE. 
Figure \ref{fig:keckdist} shows the ESI instrumental shear pattern
obtained using this method, averaging over 20 fields in order to
overcome noise, as the error on the shear is 0.09 in a $1'$ square bin
on a given field. We find after this averaging that the shear has a
mean of $0.2\%$ and is $<0.3$\% everywhere; the shear
measured fluctuates by $<0.1$\% as we average over different sets of
fields. This implies that, since our results deal with shears of
$\sim$ 1\%, and since these intrinsic values add in quadrature to
negligibly affect these shears, we can neglect this effect in our
analysis.  (Note that the ESI field has an illuminated area which is
rotated from the $x-y$ axes by $7^\circ$, accounting for the slightly
slanted geometry on this figure.) The magnification and rotation
components were also found to be $<0.3$\%, and are consequently
negligible over such a small field.

\subsection{Point Spread Function} \label{psf}

We now turn to more important sources of systematic error which need
correction: the isotropic smearing of images due to the atmosphere and
telescope optics, together with anisotropic smearing due to tracking
errors, dither alignment and optics.

We used Kaiser's {\tt imcat} software (Kaiser et al 1995) to detect
objects on our images, and measure ellipticity, radius, magnitude and
polarisability (responsiveness to image shear/smear; a large galaxy
will be less affected by a given smear than a small galaxy) for each
object. Our procedure is described in detail in BRE.


\begin{figure} \psfig{figure=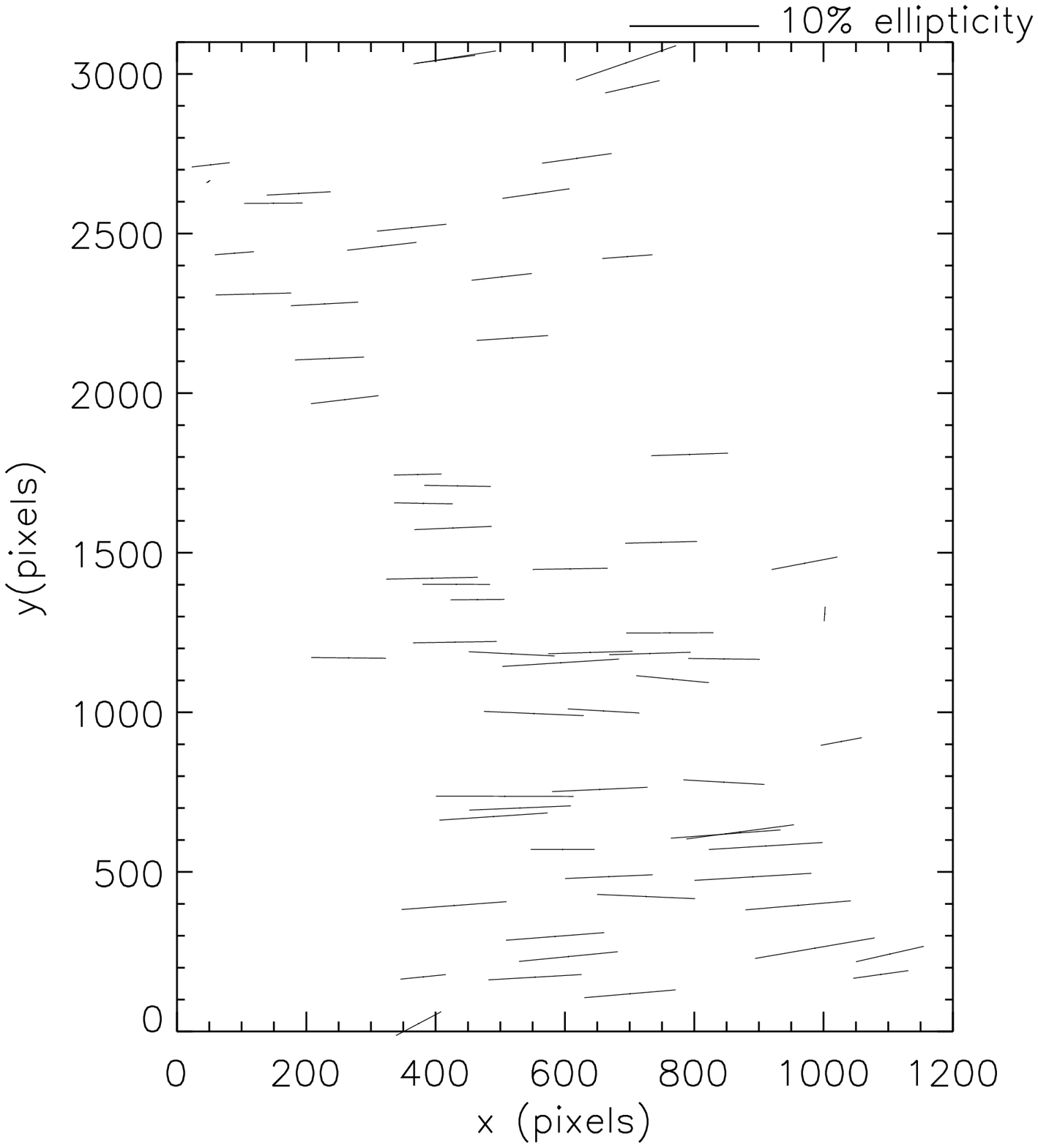,width=40mm,height=100mm}
\psfig{figure=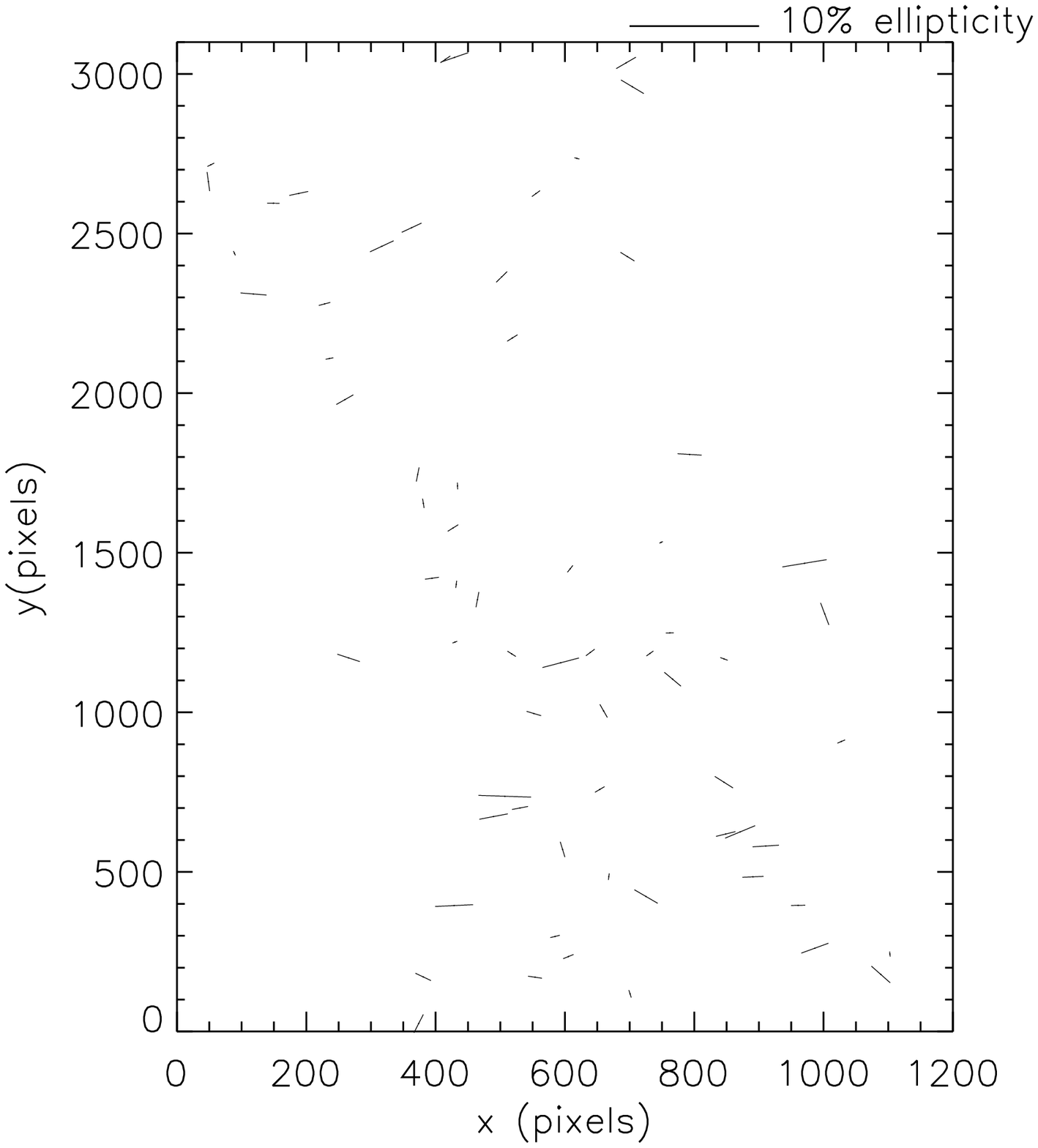,width=40mm,height=100mm} \caption{Example
stellar ellipticity pattern for ESI. (Left) Before correction; the
pattern varies qualitatively between fields but mean ellipticity here
is typical of the field-to-field average of 0.035. (Right) After
correction; mean ellipticity reduced to $<$0.001 across the survey,
with residual orientations essentially randomised.}  \label{fig:kecke}
\end{figure}

Noisy detections in our {\tt imcat hfindpeaks} catalogues were
removed using criteria as in BRE ($r_g > 1.0, \nu > 15, e < 0.5$).
Note that after this, all objects are weighted equally.  The
stars to be used to monitor the PSF were selected from the
non-saturated locus on a magnitude-radius plane for each field. In
most observations, the distribution of stellar ellipticities over the
field is found to be smooth and slowly varying (see figure
\ref{fig:kecke} for an example). Some exposures taken during July
2000 were found to have an unexplained, sharp discontinuity in this
pattern $\frac{2}{3}$ of the way up the field.  For these fields,
only objects on one side of the PSF discontinuity were used. The rms
field-to-field stellar ellipticity over all used Keck data was found
to be $\sigma_{e*}\simeq 0.035$; and reduced to a negligible $<$0.001
by our analysis. 

We then measured shear estimators for each galaxy and corrected them
for convolution with the local PSF as described in BRE. To
interpolate the PSF to the position of each galaxy, we iteratively
spatially fitted the measured stellar ellipticities with a 2-D
polynomial, removing extreme outliers due to noise and blended
images.  Similarly, each component of the smear polarisability tensor
$P^*_{sm}$ was fitted with a 2-D polynomial.  Although individually
smooth, the PSF patterns were found to have large variations from
field to field. The degree of fitting polynomial was thus adapted to
suit each pattern. Typically, a quadratic or cubic component was
necessary in the $y$ direction; the much more narrow $x$ direction
generally required only a constant or linear function. We avoided
high-order polynomial PSF models which diverged towards the edges of
the field and would have spuriously elongated galaxies by
over-correcting for the true PSF smearing effect (see Massey et al.
2001).


\begin{figure} \psfig{figure=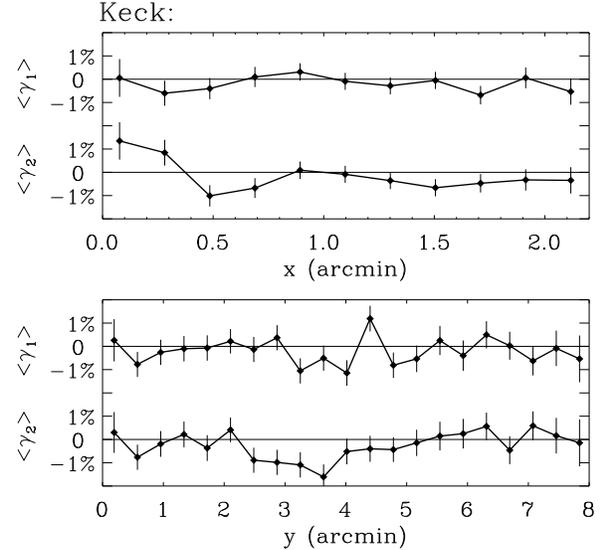,width=80mm} \caption{Average
shear of all galaxies in all Keck fields as a function of position on
chip.  Overall, $\langle\gamma_1\rangle~= -0.02\% \pm 0.16\%$ and
$\langle\gamma_2\rangle~= -0.29\% \pm 0.16\%.$} \label{fig:keck_xg}
\end{figure}


\subsection{Masking of the Field} \label{keckmask}

As a check for systematic effects associated with data reduction or
chip behaviour, we took shear estimators from all galaxies in all
fields and averaged as a function of CCD position (see figure
\ref{fig:keck_xg}). We find that the mean shear for our entire
ensemble is $\gamma_1 = -0.0002 \pm 0.0016$ and $\gamma_2 = -0.0029
\pm 0.0016$. This is consistent with zero offset in the whole
ensemble, as we would expect. Figure \ref{fig:keck_xg} further
demonstrates that there is no significant structure in our shear
values with position on the chip.

This plot proved very helpful to expose problems with the different
parts of the analysis (see discussion in Massey et al. 2001) . In
particular, the necessity for stringent masking of the field (see
also \S\ref{keckdr}) quickly became manifest. The plot not only
mirrored the locations of obvious CCD defects, but also indicated
which other regions of the chip were unsuitable for high-precision
shape measurements. In particular, galaxies at the edges of the
data/mask are cut apart by the image and appeared aligned with the
boundary. This caused a 2\% mean shear offset inside ESI's long and
thin field geometry. Other galaxies also appeared spuriously aligned
in regions where flat-fielding of any of the three dithered exposures
was imperfect (e.g. near the edge of any one of the dithers; near
internal reflections of bright stars outside the field of view; or
through cirrus). In these instances, the dithers have slightly
different background levels and image co-addition leads to residual
ellipticities in galaxies around these locations. If they are not
excluded, the mean shear throughout the field is again offset by
$\sim$1\%. we found that the masking of approximately one quarter of
the CCD reduced these effects to negligible levels in $\sim$12.8
square arcminutes of selected data per field.

\vspace{5mm}

\section{WHT data} \label{wht}

\subsection{Observations}

The WHT data used includes the 13 $8'\times16'$ fields used in our
previous detection of cosmic shear (BRE), as well as 7
fields observed in June 2000 with the new larger field of view for the
Prime Focus Camera ($16' \times 16'$). Thus the total area of the
combined WHT survey is 1.0 square degree.

The fields were observed using the Harris $R$ filter with the William
Herschel Telescope Prime Focus Camera. The $16' \times 16'$ field has
a pixel size of $0.24''$; for each field four dithers were observed,
each offset by $5''$ in order to estimate telescope shear estimation
and to remove cosmetic defects. As for the Keck survey, we observed
the fields as they passed through meridian to minimise flexure-induced
distortions. Bias frames, sky flats and standard stars were acquired
each night, and refocusing was carried out several times per night.

The median seeing for the WHT fields is $0.8''$, which is adequate
for our purposes (see Figure 8 of Bacon et al 2001); Figure
\ref{fig:seeing} records the seeing values for all fields. No data
was used with seeing $> 1.0''$.

The dithers were exposed for 900 seconds each, amounting to a 1 hour
exposure on each field. The median magnitude of detected galaxies was
$R$=25.0, with an imcat signal-to-noise of 5.0 being reached for
galaxies at $R$=25.8, similar to our previous survey fields. We keep
15 resolved galaxies per square arcminute in our final object
catalogue (see section \ref{psf} for the selection criteria), which
corresponds to a final magnitude median of $R\simeq23.5 \pm 0.2$,
including reddening corrections. As with Keck, the distribution of
magnitudes for used galaxies extends substantially fainter than the
median, with the galaxy count as a function of magnitude reaching 50\%
of maximum at $R=24.4$. Figure \ref{fig:depth} records the
reddening-corrected median magnitude vs number densities for all
fields.  According to e.g. Cohen et al (2000), the above median
magnitude corresponds to a median source redshift of $z_s \simeq 0.8$;
using a similar estimate for the redshift error to that in section
3.1, we find an uncertainty in this quantity of 0.06. As this may be
subject to sample variance, we will conservatively use a rounded
uncertainty of 0.1.

\subsection{Data Reduction} \label{whtdr}

The reduction of the WHT fields followed the same standard procedure
as outlined in BRE and \S\ref{keckdr} above. Bias
subtraction and flat fielding were followed by the elimination of
fringing, which occurs on the fields at 0.5\% of sky background
level. In order to defringe, all science exposures for each night were
stacked without offsetting, using sigma-clipping to remove objects;
this provided a fringe frame for each night. A multiple of the fringe
frame which minimised rms background on each dither was subtracted,
resulting in the fringes being reduced to an $<0.05$\% level, as in
BRE. Astrometric matching and stacking of dithers then proceeded
in the same manner as for the Keck data.

\subsection{Instrumental Distortions}

\begin{figure} \psfig{figure=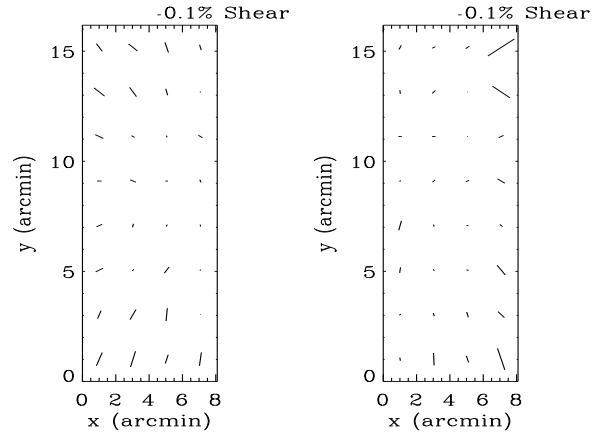,width=80mm,height=60mm}
\caption{Example of astrometric distortions for both chips of the Wide
Field Camera on WHT.} \label{fig:whtdist} \end{figure}

We re-checked the instrumental distortion for the WHT fields using
the same method as for the Keck fields (see \S\ref{keckdist} and BRE).
However, the larger number of objects in common in the WHT exposures
affords a determination of the instrumental shear on individual
fields rather than having to stack many fields as we did in
\ref{keckdist}. Figure \ref{fig:whtdist} shows the WHT instrumental
shear pattern for a typical field, including both chips of the new
$16' \times 16'$ mosaic field. Since the shear has a mean of $0.1\%$
and is $<0.4$\% at the edges of the field, we again find that the
correlation function of the telescope shear is negligible and need
not be corrected for. The telescope shear estimates fluctuate by
$<0.1$\% from field to field. Magnification and rotation components
are also found to be $<0.4$\%, and are therefore negligible.

\subsection{Point Spread Function} \label{whtpsf}

Correction for isotropic and anisotropic smear components proceeded in
an entirely analogous fashion to that for the Keck data, using {\tt
imcat}'s ellipticities and polarisabilities to correct the galaxy
shear estimates (see BRE for full details). The same criteria were
used for removing noisy detections ($r_g > 1.0, \nu > 15, e >
0.5$). The rms ellipticity of our stars from field to field was
$\sigma_{e*} \simeq 0.05$ (see BRE figure 7 for an example stellar
ellipticity field).

\begin{figure} \psfig{figure=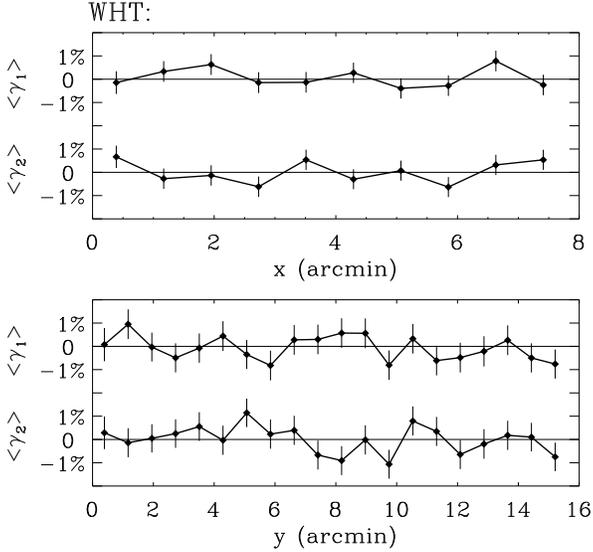,width=80mm} \caption{Average
shear of all galaxies in all WHT fields as a function of position on
chip.  Overall, $\langle\gamma_1\rangle~= 0.00\% \pm 0.13\%$ and
$\langle\gamma_2\rangle~= 0.07\% \pm 0.13\%.$} \label{fig:wht_xg}
\end{figure}

As with the Keck data, we checked for systematics in our shear
estimators by taking all shear estimators from all fields and
averaging them as a function of position (Figure \ref{fig:wht_xg}).
For WHT we find that the mean shear for our entire ensemble is
$\gamma_1 = 0.0000 \pm 0.0013$ and $\gamma_2 = 0.0007 \pm 0.0013$. This
is entirely consistent with zero offset in the whole ensemble, as we
would expect. Similarly, the figures demonstrate that there is no
significant dependence of our shear values on position on the chip.

\section{Cosmic Shear Statistics: The Correlation Function}
\label{cosshr}

We are now in possession of good quality, wide area data for cosmic
shear estimation from two telescopes. How are we to best extract the
information contained from our shear estimator catalogues, in order to
estimate cosmological parameters and the amplitude of the mass power
spectrum?

In this paper we choose to use the shear correlation function (e.g.
Kaiser 1998; Kamionkowski et al 1998), a useful probe of the shear
power on various scales. It has the advantage of being simply related
to the shear power spectrum by a Fourier transform, and affords
precise checks of the contribution to our results of PSF anisotropy
systematics. In this section, we shall describe the correlation
functions and their use as estimators for the shear signal.

The material which we have to work with is a set of catalogues
containing the positions of galaxies, plus the shear estimator
associated with each one; we therefore have a set of noisy samples of
the cosmic shear field $\gamma_{i}({\mathbf \theta})$. For this field,
we can define shear correlation functions in the following way:

\begin{eqnarray}
C_{1}(\theta) & \equiv & \langle \gamma_{1}^{r}(0)
  \gamma_{1}^{r}(\theta) \rangle \nonumber \\ 
C_{2}(\theta) & \equiv &
  \langle \gamma_{2}^r(0) \gamma_{2}^r(\theta) \rangle,
\end{eqnarray}
where the average is to be taken over all galaxy pairs separated by an
angle $\theta$. In practice, of course, we will average over an
annulus $[\theta,\theta+\Delta\theta]$.  Note that all objects that
pass the initial selection cuts (see \S\ref{keckmask}) are given
equal weight.  The shears used are those for
a rotated coordinate system, where the $x$-axis is defined by the line
joining the two galaxies. This can be achieved using the
transformations
\begin{eqnarray}
\gamma_1^r & = & \gamma_1 \cos(2 \phi) + \gamma_2 \sin(2 \phi)
\nonumber \\
\gamma_2^r & = & -\gamma_1 \sin(2 \phi) + \gamma_2 \cos(2 \phi),
\end{eqnarray}
where $\gamma_1^r$ and $\gamma_2^r$ are the rotated shears that we
require for our correlation functions. We can define a third
correlation function
\begin{equation}
C_{3} \equiv  \langle \gamma_{1}^{r}(0)
  \gamma_{2}^{r}(\theta) \rangle = \langle \gamma_{2}^{r}(0)
  \gamma_{1}^{r}(\theta) \rangle,
\end{equation}
which is expected to average to zero: under reflections $C_3$ changes
sign, so we expect equal contributions from positive and negative
shear-shear configurations. Therefore $C_3$ provides a check on
systematic effects introduced by our corrections, which need not be
parity invariant.

Finally, it is convenient to define a total correlation function
\begin{equation}
C(\theta) \equiv C_{1}(\theta) + C_{2}(\theta).
\end{equation}

The shear correlation functions can be readily calculated for any
cosmological model using
\begin{equation}
C_{i}(\theta) = \frac{1}{4\pi} \int_{0}^{\infty} dl~lC_{l}^{\gamma}
\left[ J_{0}(l\theta) + (-1)^{i+1} J_{4}(l\theta) \right],
\end{equation}
for $i=1,2$, and where $C_{l}^{\gamma}$ is the shear power spectrum
defined as in BRE (denoted as $C_{l}$ in that paper). The correlation
functions for a $\Lambda$CDM model with $\Omega_{m}=0.3$,
$\Gamma=0.21$, $\sigma_{8}=1$ and a median source redshift of
$z_{s}=1.0$ and $0.8$ are plotted in figure~\ref{fig:cth_obs}.

Since the fields are widely separated, we will not try to take the
correlation functions for the whole area represented.  Instead, we
measure the correlation functions in each separate field. However, we
need to bear in mind that the measured correlation function in a given
field is noisy due to the shot noise of the galaxy ellipticities. Thus
we require an overall estimator of the correlation function of the
cosmic shear for all fields. 

For this purpose, let us denote the correlation function measured in
field $f$ by $C_{i}^{f}(\theta)$. An estimator for $C_{i}(\theta)$ is
simply its average over all fields, i.e.
\begin{equation}
C_{i}(\theta) \simeq \frac{1}{N_{f}} \sum_{f=1}^{N_{f}}
C_{i}^{f}(\theta).
\end{equation}
Similarly, the uncertainty on our estimate of $C_{i}(\theta)$ is
simply the error in the mean given by
\begin{equation}
\sigma^{2}[C_{i}(\theta)] \simeq \frac{1}{N_{f}^{2}} 
\sum_{f=1}^{N_{f}} \left[ C_{i}^{f}(\theta) - C_{i}(\theta)
\right]^{2}.
\label{eq:ci_err}
\end{equation}
The values of $C_{i}(\theta)$ at different $\theta$ are generally not
independent. As a result, it is important for fair error estimation on
cosmological parameters to have knowledge of the covariance of the
correlation functions at different angular bins,
cov$[C_{i}(\theta),C_{j}(\theta')]$. This can easily be calculated
from the measured correlation functions using the relation
\begin{eqnarray}
&& {\rm cov}[C_{i}(\theta),C_{j}(\theta')] \simeq \nonumber \\
&& \,\,\,\,\,\,\,\, \frac{1}{N_{f}^{2}} 
\sum_{f=1}^{N_{f}} \left[ C_{i}^{f}(\theta) - C_{j}(\theta) \right] 
\left[ C_{i}^{f}(\theta') - C_{j}(\theta') \right].
\label{eq:covariance}
\end{eqnarray}

The above expression has the advantage of containing information
regarding the contribution to our errors of both shot noise and sample
variance; that is to say, it contains a measure of the entire error
budget for all scales, apart from systematic contributions. It also
allows us to account for the covariance between different angular
scales, as well as that between the different correlation functions.
We can use these estimators for the correlation function, together
with the covariance matrix, to find the best fit of cosmological
parameters to our data.

To assess the significance of the detection of the lensing signal, it
is useful to consider the errors upon the correlation functions
arising from statistical noise alone, i.e. neglecting sample
variance. It is easy to show that the covariance matrix in this case
is given by
\begin{equation}
{\rm cov}_{\rm stat}[C_{i}(\theta),C_{j}(\theta')] \simeq
\frac{\sigma_{\epsilon}^4}{N_{\rm pairs}(\theta)} \delta_{ij}
\delta_{\theta \theta'},
\label{eq:cov_stat}
\end{equation}
where $\sigma_{\epsilon}^{2} \equiv \langle \gamma_{1}^{2} \rangle =
\langle \gamma_{2}^{2} \rangle$ is the intrinsic ellipticity variance
of individual galaxies, and $N_{\rm pairs}(\theta)$ is the number of
galaxy pairs used in the angular bin centered on $\theta$.  In using
this equation, we will use the measured ellipticity dispersions of
$\sigma_{\epsilon}\simeq 0.31$ for both WHT and Keck. Note that the
covariance ${\rm cov}_{\rm stat}$ vanishes for different correlation
functions ($i\neq j$) and for different angular scales ($\theta \neq
\theta'$).

\subsection{Star-galaxy Correlation Functions}
\label{stargal}

We are able to apply the correlation function formalism to other
quantities which we have measured besides the shear field. A
particularly useful check of systematic effects is available by
considering the extra contribution to the shear correlation function
from PSF ellipticity contamination. If we have a small addition to the
shear field due to uncorrected contributions by the PSF
ellipticity,

\begin{equation}
\gamma'_i = \gamma_i + a e_i^*,
\end{equation}
then it is clear that $\left<\gamma'_i e_i^*\right> = a \left<e_i^*
e_i^*\right>$ and $\left<\gamma_i' \gamma_i'\right> = \left<\gamma_i
\gamma_i\right> + a^2 \left<e_i^* e_i^*\right>$; from this it follows
that the uncorrected ellipticities add a component to the measured
correlation function described by

\begin{equation}
C^{\rm sys}_i = \frac{\left< \gamma_i e^*_i \right>^2}{\left<
e^*_i e^*_i \right>},
\label{eq:stargalcor}
\end{equation}
where $i=1,2$. This can be directly measured from our data to
determine the error due to PSF systematics.

\subsection{Shear Variance}

Before applying these correlation functions to our data, we briefly
examine their relationship to the cell-averaged shear variance; this
is often used to quote and compare cosmic shear results, so it is
important to be able to convert between the two estimators.

Let us consider a cell with window function $W({\mathbf \theta})$
normalised as $\int d^{2}\theta~ W({\mathbf \theta}) \equiv 1$. The
average shear within this cell is

\begin{equation}
\overline{\gamma}_{i} \equiv \int d^{2}\theta~ W(\theta) 
\gamma_{i}({\mathbf \theta}).
\end{equation}

It is easy to show that the shear variance within such a cell is
related to the (total) shear correlation function by

\begin{equation}
\sigma_{\gamma}^{2} \equiv \langle |\overline{\gamma}|^{2} \rangle
 \simeq \int d^{2}\theta~ W({\mathbf \theta}) C(\theta)
\label{eq:corshr}
\end{equation}
where we have used the small angle approximation. The shear variance
is thus simply the average of the correlation function over the area
of the aperture. It is also easy to show that the error variance in
measuring $\sigma_{\gamma}^{2}$ is related to that for the correlation
functions by

\begin{equation}
\sigma^{2}[\sigma_{\gamma}^{2}] = \int d^{2}\theta \int d^{2}\theta'~
W({\mathbf \theta}) W({\mathbf \theta}') 
{\rm cov}[C(\theta),C(\theta')].
\end{equation}

For a circular cell of radius $\theta$, we can calculate these
integrals by noting that, for a top-hat window function, $\int
d^{2}\theta' W(\theta') \rightarrow 2 \theta^{-2} \int_{0}^{\theta}~
d\theta' \theta'$. We can then use these formulae to convert simply
between correlation function and variance measurements.

\section{Results}
\label{results}

\subsection{Correlation Functions} 

\begin{figure}
\psfig{figure=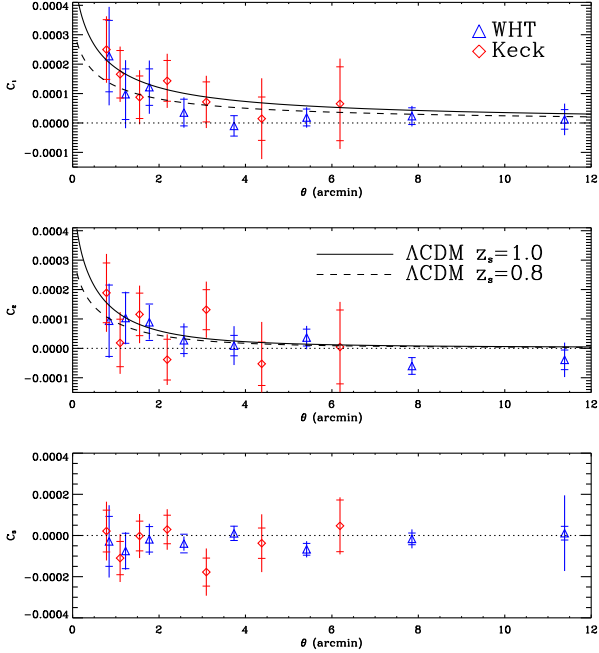,width=80mm} 
\caption{Shear correlation functions compared for the WHT and Keck. 
The inner error bars correspond to statistical errors only, while the
outer error bars correspond to the total errors (statistical + sample
variance).  For comparison, the correlation functions expected for a
$\Lambda$CDM model with $\Omega_{m}=0.3$, $\sigma_{8}=1$ and
$\Gamma=0.21$. A median source redshift of $z_{s}$=0.8 and 1.0 are
shown, as appropriate for WHT and Keck, respectively.}
\label{fig:cth_obs}
\end{figure}

Now that we have equipped ourselves with the necessary tools for
measurement, we proceed to examine the amplitude of the cosmic shear
in our data. We first measured the correlation functions defined
above, together with their covariances, for all of our cosmic shear
survey fields for Keck and WHT. Figure~\ref{fig:cth_obs} compares the
resulting shear correlation functions for both experiments, after
removal of 3$\sigma$ outliers in star-galaxy residual correlation. The
inner error bars show the statistical errors derived from
Equation~(\ref{eq:cov_stat}). The outer error bars show the total
error bar (shot noise + sample variance) derived from the diagonal
elements of the covariance matrix in Equation~(\ref{eq:covariance}).
Note that sample variance contributes significantly to our
uncertainties and should therefore be included when constraining
cosmological parameters. The off-diagonal elements (see Figures
\ref{fig:covmatrixw} and \ref{fig:covmatrixk}) allow us to quantify
the correlation between the different angular bins and between the
different correlation functions.

The expected correlation function in a $\Lambda$CDM model with
$\Omega_{m}=0.3$ is also shown in Figure~\ref{fig:cth_obs}. The shape
parameter was set to $\Gamma=0.21$, close to the values indicated by
recent galaxy surveys (Percival et al. 2001; Szalay et al. 2001) while
maintaining the $\Gamma \simeq \Omega_{m} h$ relation with
$h=0.7$. The theoretical curves were normalised by taking
$\sigma_{8}=1$, consistent with `old' cluster normalisations (Eke et
al. 1998; Viana \& Liddle 1999; Pierpaoli et al. 2001). A comparison
with the newer cluster normalisation of $\sigma_{8}=0.7$ (Borgani et
al 2001; Seljak 2001; Reiprich \& B\"{o}hringer 2001; Viana et
al. 2001) will be presented below.  The median redshift for the
galaxies was assumed to be $z_{s}=0.8$ and $1.0$, as relevant for WHT
and Keck respectively (see \S{\ref{keck}} and \S{\ref{wht}}). For the
models, the redshift distribution of the galaxies was taken to be as
in BRE.

Importantly, despite different strategies and instruments, the WHT and
Keck measurements of $C_{1}$ and $C_{2}$ are in good agreement,
especially after taking account of the difference in
$z_{s}$. Moreover, both agree with the displayed $\Lambda$CDM models
(see below for detailed discussion). Our measurement of $C_{3}$ is
shown on the bottom panel and is consistent with zero as expected from
parity symmetry.

\begin{figure}
\psfig{figure=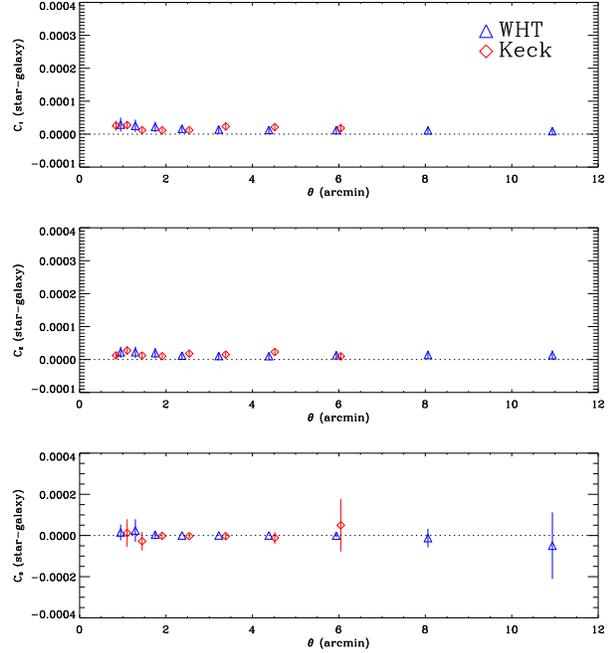,width=80mm} 
\caption{Star-galaxy correlation contribution to the error budget.
Demonstrates the negligbly small residual alignment of galaxy shear
measurements to the orientation of the original PSF (note the same
scale on the $y$-axis as figure \ref{fig:cth_obs} above).}
\label{fig:stargal} \end{figure}

We can use the formalism of section \ref{stargal} to evaluate the
systematic contribution of residual stellar anisotropy to our shear
correlation function. For this purpose, we measure the star-galaxy
correlation function given by equation (\ref{eq:stargalcor}) for all
of our WHT fields, and show the results on Figure
\ref{fig:stargal}. Note that, bearing in mind that the scale on this
figure is the same as for our shear correlation function, the
systematic contribution due to our anisotropy correction is entirely
negligible; it is lower than the shear signal by a factor $>10$
everywhere.

\subsection{Sources of Error and Covariance between Angular Scales}

\begin{figure} \psfig{figure=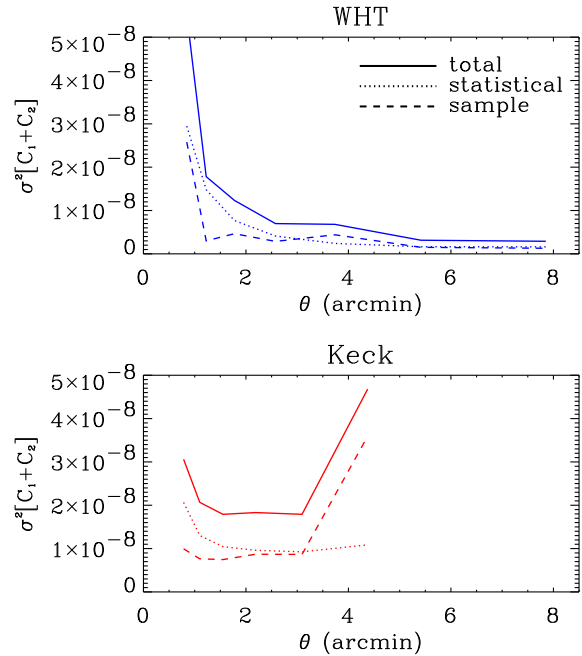,width=80mm} 
\caption{Errors in the shear correlation correlation function
$C_{1}+C_{2}$ as a function of angular scale $\theta$ for both WHT
(top) and Keck (bottom).  The total error variance (solid line) is
decomposed into a statistical (dotted line) and a sample variance
(dashed line) contribution in each case. Note that the area of the WHT
survey is a factor of about 1.7 larger than that for Keck.}
\label{fig:cth_err}
\end{figure}

It is instructive to compare the different sources of error for each
telescope, given the complementary survey strategies.
Figure~\ref{fig:cth_err} shows the error variance of the shear
correlation correlation functions $C_{1}+C_{2}$ of
Figure~\ref{fig:cth_obs} as a function of angular scale $\theta$. The
total error variance (from Eq.~[\ref{eq:ci_err}]) was decomposed into a
statistical error (from Eq.~[\ref{eq:cov_stat}]) and a sample variance
contribution (computed by subtracting the latter from the former).
The cosmic variance contribution is clearly significant for both Keck
and WHT and must be taken into account in the determination of
cosmological parameters from cosmic shear data.

The advantage of using a 10m-class telescope in reducing statistical
errors is apparent in this Figure. Indeed, the statistical errors for
Keck for $\theta \la 3'$ are significantly smaller than that for WHT,
after scaling them down by a factor of 1.7 corresponding to a common
survey area of 1 deg$^{2}$ in both surveys. The much shorter exposure
times allowed by Keck yield an improvement in the seeing, tracking
errors and overall image quality. Aided by the finer pixellisation of
the ESI CCD, this results in an increased galaxy number density and
correspondingly lower statistical errors when normalised by survey
area (as predicted by Bacon et al 2001). Of course, the statistical
errors for WHT are smaller on large angular scales ($\theta \ga 3'$)
thanks to the larger field of view of this telescope.

\begin{figure} \psfig{figure=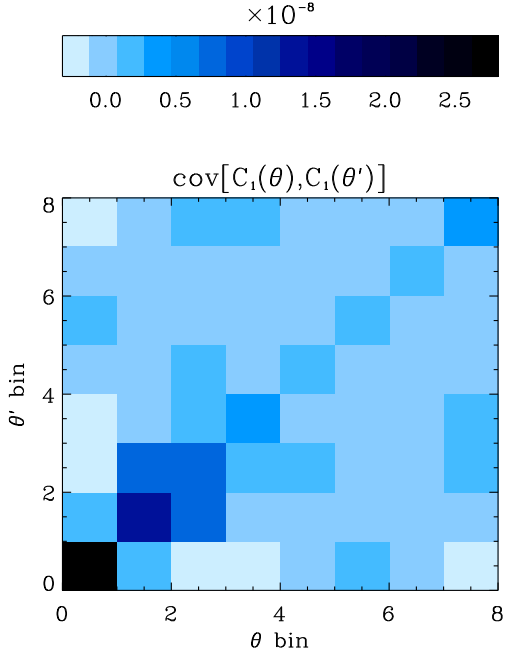,width=80mm}  
\caption{Covariance matrix of the shear $C_1$ correlation function
between the different angular bins of the WHT data plotted in figure
\ref{fig:cth_obs}.} \label{fig:covmatrixw} \end{figure}

\begin{figure} \psfig{figure=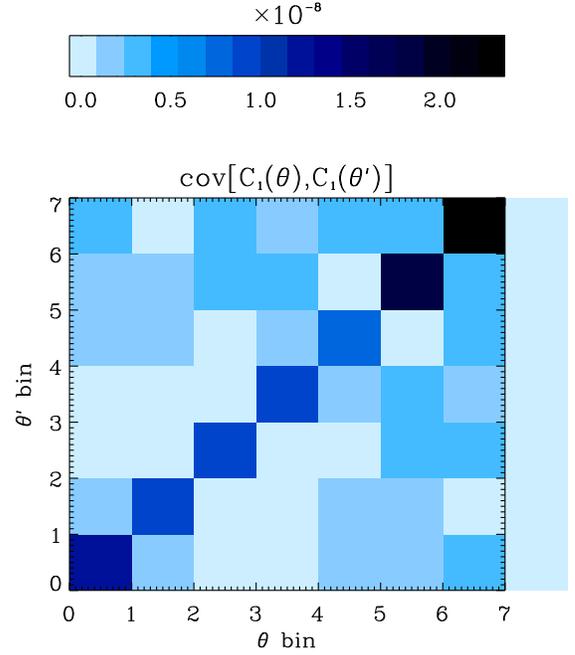,width=80mm}  
\caption{Covariance matrix of the shear $C_1$ correlation function
between the different angular bins of the Keck data plotted in figure
\ref{fig:cth_obs}.} \label{fig:covmatrixk} \end{figure}

The covariance of the shear correlation functions on different angular
scales is also non-negligible. Figures \ref{fig:covmatrixw} and
\ref{fig:covmatrixk} show the covariance matrix for the $C_{1}$
correlation function for WHT and Keck, respectively. In each case, the
angular bins correspond to the $\theta$ values shown in
Figure~\ref{fig:cth_obs}. The WHT correlation function clearly
has important covariance between bins on small angular scales ($\theta
\la 3'$). The Keck correlation functions also has important covariance
on large scales ($\theta \ga 3'$). This is due to the elongated
field geometry of the Keck camera. In both cases, it is important
to include the full covariance matrix for the determination of
cosmological parameters.

\subsection{Shear Variance}

We now wish to compare our results to those obtained recently by
other groups. Since many of these results are quoted in terms of the
angular dependence of the shear variance, we use equation
(\ref{eq:corshr}) to convert our correlation function measurements
into shear variance measures as a function of scale.

\begin{figure}
\psfig{figure=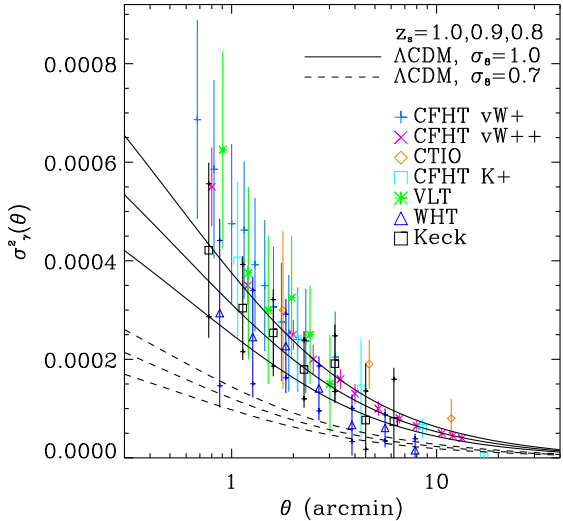,width=80mm} 
\caption{Shear variance as a function of aperture radius. Our
measurements with Keck and WHT are shown along with those of other
groups (see text). Again, the inner and outer error bars correspond to
statistical and total uncertainties, respectively.  The expected shear
variances for a $\Lambda$CDM model with $\Omega_{m}=0.3$, and
$\Gamma=0.21$ are also shown for a range of values for the median
source redshift $z_{s}$ corresponding approximately to the uncertainty
in this parameter. The models are shown both for $\sigma_{8}\simeq 1$
(solid lines) and $\sigma_{8}\simeq 0.7$ (dashed) cluster
normalisation.}
\label{fig:var_obs}
\end{figure}

The results for the WHT and Keck surveys, are plotted in
Figure~\ref{fig:var_obs}. Taking the most significant point in each
case and using statistical errors only, we find that the cosmic shear
signal is detected at the $3.7\sigma$, $3.5\sigma$ and $5.1\sigma$
level with Keck, WHT, and both combined, respectively. Also plotted
are the predictions for the $\Lambda$CDM model with $\Omega_{m}=0.3$
and $\Gamma=0.21$ as before. They are shown for a range of values of
the galaxy median redshift $z_{s}$, corresponding roughly to the
uncertainty and survey-to-survey variations in this parameter. Both
old ($\sigma_{8}=1$) and new cluster normalisations
($\sigma_{8}=0.7$) are shown (see discussion below). We also plotted
the measurements from other groups, namely from van Waerbeke et al
2000 (CFHT vW+), Kaiser et al. 2000 (CFHT K+), Wittman et al 2000
(CTIO), Maoli et al 2001 (VLT), and van Waerbeke et al 2001 (CFHT
vW++). Note that the errors for the VLT, CFHT vW+ and CFHT vW++ do
not include cosmic variance. All of the above groups are obtaining
results which are broadly consistent with each other and with
$\Lambda$CDM with the older normalisation of $\sigma_8 = 1$.

\subsection{Cosmological Constraints}
Now that we have measurements of cosmic shear on several scales, we
can determine the implications for cosmological parameters.  In order
to do this, we use a Maximum Likelihood approach. We first construct
data vectors ${\mathbf d}\equiv \{ C_{1}(\theta_{n}),
C_{2}(\theta_{n}) \}$ which are simply a rearrangement of our observed
correlation functions. We consider a $\Lambda$CDM model with two
parameters, $a_{1}=\Omega_{m}$ and $a_{2}=\sigma_{8}$. The shape
parameter $\Gamma$ for the matter power spectrum is set to 0.21 as
indicated by recent measurements of galaxy clustering (Percival et
al. 2001; Szalay et al. 2001). Using the formalism described in BRE
and in section \ref{cosshr}, we can compute the correlation
function for any values of these parameters and arrange them as a
theory vector ${\mathbf t}({\mathbf a})$ set out in exactly the same
form as our data vector. The median redshift $z_{s}$ is set to 0.8 and
1.0 for WHT and Keck respectively.

Because the correlation functions were derived from an average over a
large number of fields, the central limit theorem ensures that our
errors (and covariances) are gaussian. In this case the log-likelihood
is simply
\begin{equation}
\chi^{2}=[{\mathbf d}-{\mathbf t}(a)]^{T} {\mathbf V}^{-1}[{\mathbf
d}-{\mathbf t}(a)],
\end{equation}
where ${\mathbf V}$ is the covariance matrix computed using
Equation~(\ref{eq:covariance}). We minimise this quantity as a
function of ${\mathbf a}$ to find the best estimate of the
normalisation of the power spectrum. To compute confidence contours,
we numerically integrate the probability distribution function
$P({\mathbf a}) \propto e^{-\chi^2/2}$. 

\begin{figure}
\center
\psfig{figure=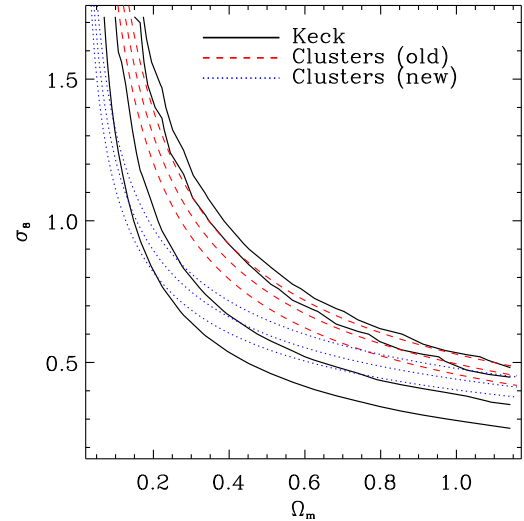,width=80mm} 
\caption{Constraints on the joint distribution of $\Omega_m$ and
$\sigma_8$ from our Keck measurement. The solid
contours correspond to 68\% and 95\% confidence levels. Note that these
contours include statistical errors and non-gaussian sample variance,
but do not include the uncertainty in the galaxy redshifts and in the
shear measurement method. Also
shown are the $1\sigma$ contours from the old (Pierpaoli et al. 2001)
and new (Seljak 2001) cluster abundance normalisation.}
\label{fig:chi2_keck}
\end{figure}

\begin{figure}
\center
\psfig{figure=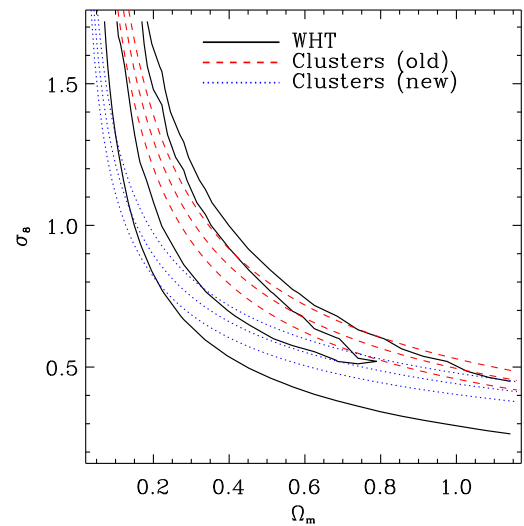,width=80mm} 
\caption{Constraints on the joint distribution of $\Omega_m$ and
$\sigma_8$ as in the previous figure, but for our WHT measurement.}
\label{fig:chi2_wht}
\end{figure}

The constraints for our Keck and WHT data, taken separately, are shown
on figures~\ref{fig:chi2_keck} and \ref{fig:chi2_wht}. Note that they
do not include the uncertainty in $z_{s}$ and in the shear measurement
method. These two sources of errors will be included in our final
estimate below. The constraints from each of the data set are
consistent with each other. As is apparent on the figures, the
constraints reveal the well known degeneracy between $\sigma_{8}$ and
$\Omega_{m}$ when 2-point statistics alone are used. The wider angular
range of WHT allow us to start breaking this degeneracy, by rejecting
large values of $\Omega_{m}$ (see van Waerbeke et al. 2001). Note that
the width of the contours for Keck and WHT are comparable, even though
the WHT area is larger by about 70\%. This results from the larger
number density of galaxies afforded by Keck thanks to the better
seeing statistics and smaller pixel size (see discussion in Bacon et
al. 2001).

\begin{figure}
\center
\psfig{figure=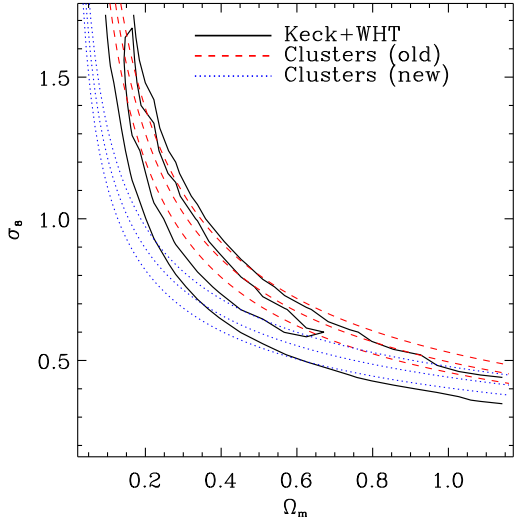,width=80mm} 
\caption{Constraints on the joint distribution of $\Omega_m$ and
$\sigma_8$ for the combination of our Keck
and WHT measurements.}
\label{fig:chi2_comb}
\end{figure}

The constraints on $\sigma_{8}$ and $\Omega_{m}$ obtained by combining
the Keck and WHT data are shown in Figure~\ref{fig:chi2_comb}. As
expected, the combined contours are consistent with the two
measurements taken separately. A good fit to our 68\% confidence level
is given by
\begin{equation}
\sigma_{8} \left( \frac{\Omega_{m}}{0.3} \right)^{0.68} =
0.97^{+0.10}_{-0.09} ~~~~{\rm (shot\ noise + sample\ variance)}
\end{equation}
with $0.14<\Omega_m<0.65$. To this statistical error (which includes
non-gaussian sample variance), we also must add that arising from the
uncertainty the in the redshift distribution and in the method
calibration.  A good estimate for these uncertainties can be obtained
by noting that the shear variance on the central scale for our
experiments, i.e. circular radius of 3 arcmins, scales as
$\sigma^{2}_{\gamma} \propto P_{\gamma}^{-2} \sigma_{8}^{2.5}
z_{s}^{1.6}$, where $P_{\gamma}^{-1}$ is the ellipticity-to-shear
conversion factor in the KSB method (see BRE). This factor was shown
to have a $1\sigma$ uncertainty of 5\% (BRE). The
uncertainty in $z_{s}$ is about 12\% for our two data sets (see
\S\ref{keck} and \S\ref{wht}). Using the former expression to
propagate the errors yields a final constraint on the amplitude of the
matter power spectrum of $\sigma_{8} (\Omega_{m}/0.3)^{0.68} =
0.97^{+0.10}_{-0.09} \pm 0.07 \pm 0.04$, where the error bars denote
68\% confidence levels, arising from statistical (incl. sample
variance), redshift uncertainty and $P_{\gamma}$ uncertainty,
respectively. Adding these errors in quadrature, we obtain our final
estimate for the amplitude of the power spectrum, namely
\begin{equation}
\sigma_{8} \left( \frac{\Omega_{m}}{0.3} \right)^{0.68} = 0.97 \pm 0.13~~~{\rm (total~error)},
\end{equation}
where the error is the total error for a 68\% confidence level.

Our results are consistent with the results of van Waerbeke et
al. (2002) who found $\sigma_{8}=(0.94\pm0.07)
(\Omega_m/0.3)^{0.24^{-0.18}_{+0.18}\Omega_{m} - 0.49}$ (68\%CL),
after marginalising over $\Gamma$ in the range 0.1 to 0.4. They are
also consistent with those of Hoekstra et al. (2002) who found
$\sigma_{8}=0.81^{+0.14}_{-0.19}$ (95\%CL) for $\Omega_{m}=0.3$ and
$\Gamma=0.21$.

It is interesting to compare our measurement of $\sigma_{8}$ to that
derived from cluster abundance. Our results are in agreement with the
older normalisation of the power spectrum from cluster abundance.
Until recently, the cluster abundance was indeed quoted to be
$\sigma_{8}=(1.00\pm0.07) (0.3/\Omega_{m})^{0.60}
(\Gamma/0.21)^{0.28-0.31 \Omega_{m}}$ ($1\sigma$; Pierpaoli et al.
2001; see also earlier and consistent estimates by Eke et al. 1998;
Viana and Liddle 1999). Recently the cluster normalisation has been
revised to a lower value, mainly because of the use of the observed
mass-temperature relation for clusters, rather than the simulated one.
For instance, Seljak (2001) finds $\sigma_{8} = (0.75 \pm 0.06)
(0.3/\Omega_{m})^{0.44} (\Gamma/0.21)^{0.08}$ ($1\sigma$; see also
Borgani et al 2001; Rieprich \& B\"{o}hringer 2001; Viana et
al. 2001). A similar normalisation was found by combining the 2dF
galaxy survey with CMB measurements by Lahav et al. (2001). The two
cluster normalisations are also plotted on
Figures~\ref{fig:chi2_keck}, \ref{fig:chi2_wht} and
\ref{fig:chi2_comb}. Clearly our results are most consistent with a
$\sigma_8=1$ normalisation, but do not rule out the new normalisation
which is consistent at the 1.7$\sigma$ level for $\Omega_m=0.3$.  This
can also be seen by examining Figure~\ref{fig:var_obs} which shows the
impact of these different normalisations for the shear variance.

\section{Conclusions}
\label{conclusions}

We have presented a thorough analysis of weak lensing signals obtained
using two telescopes. The independent instruments introduce different
systematic effects upon galaxy shapes, the cross-checking and control
of which is the most important challenge facing current measurements
of weak shear. Our many-line-of-sight survey strategy for
cosmic shear, using 173 Keck telescope fields, also complements an
alternative 20 large-field survey strategy with the William Herschel
Telescope.

We have measured cosmic shear at a signal-to-noise of 5.1$\sigma$ with
both Keck and WHT, and have measured the amplitude of the power
spectrum, $\sigma_{8} \left( \frac{\Omega_{m}}{0.3} \right)^{0.68} =
0.97 \pm 0.13$ with $0.14<\Omega_m<0.65$, including all contributions
to the 68\% CL uncertainty: statistical noise, sample variance,
covariance between angular bins, systematic effects, and redshift
uncertainty. A measurement of this quantity from cosmic shear is
cosmologically valuable, as it represents a direct measure of the
amplitude of mass fluctuations.

These measurements have been obtained after a careful study and
removal of systematic effects which can mimic a shear signal at the
1\% level. We have demonstrated for both telescopes that no offset in
the ensemble of shear estimators is found as a function of position
on images, and that the contribution of star-galaxy correlations is
negligible with our catalogue selection. Our methods have been tested
thoroughly in the context of detailed simulations of realistic images
(Bacon et al. 2001) and have been shown to operate successfully in
recovering shear at the necessary level. 

Our results for Keck and WHT are consistent with each other,
strengthening confidence in control of systematics. The joint results
are also consistent with other recent cosmic shear measurements
(Hoekstra et al.  2002; van Waerbeke et al. 2002). They also agree
with the old cluster abundance normalisation (Eke et al. 1998; Viana
and Liddle 1999; Pierpaoli et al. 2001). Our results prefer this
normalisation, but can not rule out lower cluster-abundance
normalisation which has recently been derived (Borgani et al 2001;
Seljak 2001; Rieprich \& B\"{o}hringer 200; Viana et al. 2001; see
also the similar normalisation derived from 2dF+CMB by Lahav et
al. 2001). This discrepancy, if confirmed, could arise from unknown
systematics in either the cluster or cosmic shear methods. Note for
instance that uncorrected systematics in cosmic shear measurements
will tend to add to the lensing signal and thus lead to an
overestimation of $\sigma_{8}$. For the cluster method, further
studies would be needed to understand the difference between the
observed mass-temperature relation and that found in numerical
simulations. It is important to understand the origin of the
discrepancy between cosmic shear and cluster abundance methods. If
this is not explained by such systematics, it could point towards a
failure of the standard $\Lambda$CDM paradigm, and therefore have
important consequences for cosmology.

\section*{Acknowledgements} We are indebted to Nick Kaiser for
providing us with the {\tt imcat} software, and to Douglas Clowe for
advice on its use. We thank Mike Bolte and staff at the Keck
Observatory for their assistance in implementing the new wide-field
filter on the Echellette Spectrographic Imager. We thank Sarah Bridle,
Tzu-Ching Chang, and Jason Rhodes for useful discussions. We also
thank Mark Sullivan for his help in securing some of the Keck data and
Max Pettini for providing us with one of the WHT fields. DJB was
supported by a PPARC postdoctoral fellowship. AR was supported by a
TMR postdoctoral fellowship from the EEC Lensing Network, and by a
Wolfson College Research Fellowship.

\bsp

\label{lastpage}


\begin{thebibliography}{99}
\bibitem{bac00} Bacon D. J., Refregier A. R., Ellis R. S., 2000, MNRAS, 318,
625 (BRE)
\bibitem{bac01a} Bacon D. J., Refregier A., Clowe D., Ellis R. S.,
2001, MNRAS, 325, 1065
\bibitem{bar99} Bartelmann M., Schneider P., 2000, Physics Reports,
340, 291
\bibitem{b2} Baugh C. M., Cole S., Frenk C. S., Lacey C. G., 1998,
ApJ, 498, 504
\bibitem{bernard} Bernardeau F., van Waerbeke L., Mellier Y.,
1997, A\&A, 322, 1
\bibitem{b3} Bertin E., Arnouts S., 1996, A\&AS, 117, 393
\bibitem{ber97} Bernardeau, F., van Waerbeke, L., \& Mellier, Y.,
  1997, A\&A, 322, 1
\bibitem{b4} Blandford R. D., Saust A. B., Brainerd T. G., Villumsen
J. V., 1991, MNRAS, 251, 600
\bibitem{bor} Borgani S., Rosati P., Tozzi P., Stanford S. A.,
Eisenhardt P. R., Lidman C., Holden B., Della Ceca R., Norman C.,
Squires G., 2001, ApJ, 561, 13.
\bibitem{brow00} Brown M. L., Taylor A. N., Hambly N. C., Dye S., 2000,
astro-ph/0009499
\bibitem{car95} Carter D., Bridges T., 1995, {\it WHT Prime Focus and
Auxiliary Port Imaging Manual}, available at
http:/lpss1.ing.iac.es/manuals/html\_manuals/wht\_instr/pfip
\bibitem{cat00} Catelan P., Kamionkowski M., Blandford R. D.,
  2001, MNRAS, 320, L7
\bibitem{coh} Cohen J. G., Hogg D. W., Blandford R., Cowie L. L., Hu E.,
Songaila A., Shopbell P., Richberg K., 2000, ApJ, 538, 29
\bibitem{crit00} Crittenden R. G., Natarajan P., Pen U.-L., Theuns T., 2000,
astro-ph/0012336
\bibitem{cro00} Croft R. A. C., Metzler C. A., 2000, ApJ, 545, 561
\bibitem{eke98} Eke V. R., Cole S., Frenk C., Henry H. J., 1998,
  MNRAS, 298, 1145
\bibitem{erb01} Erben T., van Waerbeke L., Bertin E., Mellier Y.,
  Schneider P., 2001, A\&A, 366, 717
\bibitem{ham01} H\"{a}mmerle et al., 2001, submitted to A\&A, preprint
  astro-ph/0110210
\bibitem{heav} Heavens A., Refregier A., Heymans C., 2000, MNRAS, 319,
  649
\bibitem{hoe98} Hoekstra H., Franx M., Kuijken K., Squires G., 1998,
  ApJ, 504, 636
\bibitem{hoe02} Hoekstra H., Yee, H.K.C., Gladders, M.D., Felipe
  Barrientos, L., Hall, P.B., \& Infante, L., 2002, submitted
  to ApJ, preprint astro-ph/0202285
\bibitem{b9} Jain B., Seljak U., 1997, ApJ, 484, 560
\bibitem{b10} Jain B., Seljak U., White S., 2000, ApJ, 530, 547
\bibitem{b12} Kaiser N., 1992, ApJ, 388, 272
\bibitem{b14} Kaiser N., Squires G., Broadhurst T., 1995, ApJ, 449, 460
\bibitem{b13} Kaiser N., 1998, ApJ, 498, 26
\bibitem{kai00b} Kaiser N., Wilson G., Luppino G. A., 2000, astro-ph/0003338
\bibitem{kam97} Kamionkowski M., Babul A., Cress C. M., Refregier
  A., 1998, MNRAS, 301, 1064
\bibitem{b15} Kauffmann G., Guiderdoni B., White S. D. M., 1994,
MNRAS, 267, 981
\bibitem{lah01} Lahav, O., et al. 2001, submitted to MNRAS, astro-ph/0112162
\bibitem{lup97} Luppino G. A., Kaiser N., 1997 , ApJ, 475, 20
\bibitem{lup93} Lupton R., 1993, Statistics in Theory and
Practice (Princeton U. Press: Princeton)
\bibitem{mao} Maoli R., van Waebeke L., Mellier Y., Schneider P., Jain
B., Bernardeau F., Erben T., Fort B., 2001, A\&A, 368, 766
\bibitem{mas01} Massey R., Bacon D., Refregier A., Ellis R., to appear in 
{\it A New Era In Cosmology}, eds. T. Shanks \& N. Metcalfe, ASP
Conference Series, preprint astro-ph/0112393
\bibitem{b17} Mould J., Blandford R., Villumsen J., Brainerd T., Smail
I., Smail I., Small T., Kells W., 1994, MNRAS, 271, 31
\bibitem{pie01} Pierpaoli E., Scott D., White M., 2001,
  MNRAS, 325, 77
\bibitem{per01} Percival W. et al., 2001, MNRAS, 327, 1297
\bibitem{rei01} Reiprich T.H. \& B\"{o}hringer H., 2001, to appear
  in ApJ, preprint astro-ph/0111285
\bibitem{ref02} Refregier A., Rhodes, J., Groth E., 2002, submitted
  to ApJL
\bibitem{rho99a} Rhodes J., 1999, PhD thesis, Department of Physics,
  Princeton University
\bibitem{rho99} Rhodes J., Refregier A., Groth E., 2000,
ApJ, 536, 79
\bibitem{rho99b} Rhodes J., Refregier A., Groth E., 2001, ApJ, 552, L85
\bibitem{sel01} Seljak U., 2001, submitted to MNRAS, preprint
  astro-ph/0111362
\bibitem{sch92} Schneider P., Ehlers J., Falco E. E., 1992, {\it Gravitational 
Lenses}, Springer Verlag, Heidelberg
\bibitem{sza01} Szalay A. et al., 2001, preprint astro-ph/0107419
\bibitem{van98} Van Waerbeke L., Bernardeau F., Mellier Y.,
  1999, A\&A, 342, 15
\bibitem{van00} Van Waerbeke L., Mellier Y., Erben T., Cuillandre
J. C., Bernardeau F., Maoli R., Bertin E., McCracken H. J., Le
F\`{e}vre O., Fort B., Dantel-Fort M., Jain B., Schneider P., 2000,
A\&A, 358, 30
\bibitem{van00b} Van Waerbeke L., Mellier Y., Radovich M., Bertin E.,
Dantel-Fort M., McCracken H. J., Le F\`{e}vre O., Foucaud S.,
Cuillandre J.-C., Erben T., Jain B., Schneider P., Bernardeau F., Fort
B., 2001, A\&A, 374, 757
\bibitem{van02} Van Waerbeke, L., Mellier, Y., Pell\'{o}, R., Pen,
  U.-L., McCracken, H.J., \& Jain, B., 2002, submitted to A\&A, 
  preprint astro-ph/0202503
\bibitem{b26} Viana P., Liddle A., 1999, MNRAS, 303, 535
\bibitem{via01} Viana P.T.P., Nichol R.C., Liddle A.R., 2001,
  submitted to ApJL, preprint astro-ph/0111394
\bibitem{wit00} Wittman D. M., Tyson J. A., Kirkman D., Dell'Antonio
I., Bernstein G., 2000, Nature, 405, 143
\end{thebibliography}
\end{document}